\documentclass[conference]{IEEEtran}
\IEEEoverridecommandlockouts
\usepackage{cite}
\usepackage{amsmath,amssymb,amsfonts}
\usepackage{algorithmic}
\usepackage[ruled,linesnumbered]{algorithm2e}
\usepackage{graphicx}
\usepackage{textcomp}
\usepackage{tabularx}
\usepackage{ragged2e} 
\usepackage{booktabs} 
\usepackage{multirow}
\usepackage{enumitem}

\usepackage{xcolor}
\newcolumntype{Y}{>{\centering\arraybackslash}X}
\newcolumntype{L}{>{\centering\arraybackslash}c}
\newcolumntype{C}[1]{>{\raggedright\arraybackslash}p{#1}}

\def\BibTeX{{\rm B\kern-.05em{\sc i\kern-.025em b}\kern-.08em
    T\kern-.1667em\lower.7ex\hbox{E}\kern-.125emX}}
\begin{document}

\title{MARS: Exploiting Multi-Level Parallelism for DNN Workloads on Adaptive Multi-Accelerator Systems}

\author{\IEEEauthorblockN{Guan Shen$^{1}$,
Jieru Zhao$^{1*}$, Zeke Wang$^{2}$,
Zhe Lin$^{3}$, Wenchao Ding$^{4}$, Chentao Wu$^{1}$, Quan Chen$^{1}$, Minyi Guo$^{1*}$}
\IEEEauthorblockA{${}^{1}$Shanghai Jiao Tong University, ${}^2$Zhejiang University, ${}^3$Sun Yat-sen University, ${}^4$Fudan University\\
\{shenguan,zhao-jieru\}@sjtu.edu.cn, wangzeke@zju.edu.cn, linzh235@mail.sysu.edu.cn,\\
dingwenchao@fudan.edu.cn, \{wuct, chen-quan, guo-my\}@cs.sjtu.edu.cn}\vspace{-1cm}}

\maketitle
\begingroup\renewcommand\thefootnote{*}
\footnotetext{Jieru Zhao and Minyi Guo are corresponding authors.}
\endgroup
\begin{abstract}
Along with the fast evolution of deep neural networks, the hardware system is also developing rapidly. As a promising solution achieving high scalability and low manufacturing cost, multi-accelerator systems widely exist in data centers, cloud platforms, and SoCs. Thus, a challenging problem arises in multi-accelerator systems: selecting a proper combination of accelerators from available designs and searching for efficient DNN mapping strategies. To this end, we propose MARS, a novel mapping framework that can perform computation-aware accelerator selection, and apply communication-aware sharding strategies to maximize parallelism. Experimental results show that MARS can achieve 32.2\% latency reduction on average for typical DNN workloads compared to the baseline, and 59.4\% latency reduction on heterogeneous models compared to the corresponding state-of-the-art method. 
\end{abstract}


\vspace{-0.2cm}
\section{Introduction}
\vspace{-0.1cm}
DNN models have achieved cutting-edge accuracy for a wide range of tasks in various areas like computer vision\cite{he2016deep}, natural language processing\cite{vaswani2017attention}, and recommendation systems\cite{cheng2016wide}. Simultaneously, the power of DNNs imposes a significant computational and memory burden on traditional hardware systems. For example, one of the largest language models, GPT-3, involves about $175$ billion parameters and $3.14\times 10^8$ PFLOPs, which require a large amount of computational and memory resources. More critically, the completion of model training does not mean that the work is done. The trained models should be deployed to target platforms like cloud servers and edge devices for inference. These scenarios tend to be more cost-sensitive, compared to training systems.

Multi-accelerator systems have begun to attract the attention of researchers and industry. As Moore's Law fades out of effect, semiconductor manufacturers can no longer exponentially increase the computational resources in chips. In other words, manufacturing one large chip could be expensive. However, multi-accelerator systems can achieve the same performance at a lower manufacturing cost. For example, Microsoft\cite{fowers2018configurable} has applied customized accelerators based on FPGA in the data center network. Amazon Web Services has also provided FPGA clusters (F1 instances) equipped with up to 8 FPGAs. NVIDIA’s Jetson AGX Xavier is an SoC design integrating an ARM CPU, a Volta GPU, 2 NVIDIA Deep Learning Accelerators (NVDLA), and multimedia accelerators. 

These systems can achieve high performance and scalability with the interconnection between accelerators enabling collaborations. But this still requires much engineering effort and expert knowledge because of a large design space. First, for a certain layer in a given DNN workload, various accelerator designs may demonstrate performance gaps due to different computation patterns. In an adaptive multi-accelerator system, the design of accelerators can be configured, which is natural for FPGA platforms and SoCs at the design phase. Thus, one needs to choose a proper combination of accelerator designs to accommodate these layers in the workload and achieve the best overall performance. Second, to fully utilize computational resources and relieve the memory burden, parallelism strategies, like data parallelism and model parallelism, should be applied to partition the workload into accelerators with the awareness of communication. Note that multiple parallelism strategies can work together and form a multi-level decision problem. Due to these factors, an effective mapping framework is urgently required to explore the large design space.

There have been some previous works focusing on the mapping algorithm on multi-accelerator systems. 
Zhang et al.\cite{zhang2019efficient} propose a mapping strategy based on dynamic programming to partition the DNNs into different FPGAs. 
Herald\cite{kwon2021heterogeneous} focuses on mapping multi-DNN workloads onto the heterogeneous dataflow accelerators. But it lacks communication awareness during the mapping.
H2H\cite{zhang2022h2h} provides mapping strategies with computation and communication awareness for heterogeneous accelerator systems. All of them fail to perform intra-layer parallelism to fully utilize computational resources.

\begin{figure*}[t] 
    \centering
    \includegraphics[width=0.95\textwidth]{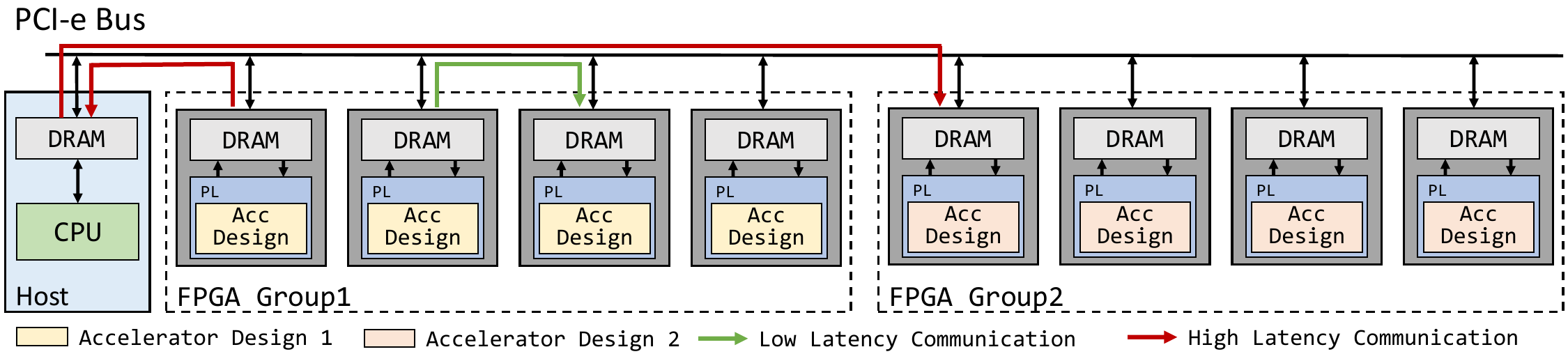}
    \vspace{-0.2cm}
    \caption{The architecture of an F1.16xlarge	instance on AWS}
    \vspace{-0.4cm}
    \label{fig:F1}
\end{figure*}

To this end, we propose MARS, a mapping framework that can exploit multi-level parallelism on adaptive multi-accelerator systems with high scalability. Our main contributions are summarized as follows:
\begin{itemize}
    \item We give a detailed system formulation together with the design space to cover current multi-accelerator systems and DNN workloads.
    \item We generalize parallelism strategies for multi-accelerator systems. We develop a general representation to describe workload partitioning, generate multi-level parallelism strategies, and integrate them for evaluation.
    \item We develop a mapping algorithm based on a two-level genetic algorithm with heuristics, which performs efficient design space exploration and finds a proper choice of accelerators, workload allocation, and parallelism strategies. 
    \item 
    MARS achieves 32.2\% latency reduction compared to the baseline for typical DNN models and outperforms the existing SOTA with 59.4\% latency reduction for mapping heterogeneous models to heterogeneous accelerators.
\end{itemize}

\vspace{-0.1cm}
\section{Background and Motivation}\label{background}
\vspace{-0.1cm}
\subsection{Adaptive Multi-Accelerator Systems} \label{Adaptive}
As more and more multi-accelerator systems are developed, adaptive multi-accelerator systems gain popularity for their flexibility. They can easily adapt to the coming workload by configuring accelerators in the system to maximize resource utilization. They can also work as a prototype validation platform to guide the design of other multi-accelerator systems. Fig. \ref{fig:F1} shows the architecture of an EC2 F1.16xlarge instance from AWS, a representative adaptive multi-accelerator system. Each instance has a host machine with eight Xilinx UltraScale+ FPGAs. The host-side applications can access/transmit data from/to FPGA's local DRAM via the PCI-e bus. The programmable logic (PL) of each FPGA can be \textbf{configured independently} through the FPGA image specified by the user. Moreover, eight FPGAs in the system are separated into two groups, as illustrated in Fig. \ref{fig:F1}. FPGAs from the same group can communicate with each other without the interference of the host, which could reduce the communication latency\cite{peer2peer}. This asymmetrical communication pattern pose new challenges. Users need to rearrange workloads to leverage the low-latency communication feature. Note that though we take the F1 instance as an example, this asymmetrical communication pattern widely exists in multi-accelerator systems in industry. Microsoft\cite{fowers2018configurable} organizes accelerators through switches from the data center network in a hierarchical manner. Accelerators in the same rack can communicate much faster compared to the pairs from different racks.

\vspace{-0.2cm}
\subsection{DNN Workloads on Multi-Accelerator Systems}

DNN workloads can be represented as a computation graph, which is a directed acyclic graph consisting of layers. For a CNN model, there are various layers including convolution, batch normalization, pooling, activation and fully-connected layers. Typically, convolution layers occupy most of the computation resources. The convolution layer can be represented by a six-level nested loop. Most CNN accelerators perform loop transformation, such as loop interchange and loop tiling to map a given convolution layer to PEs (processing elements) in hardware architectures. Accelerator designers can surely find an optimal set of parameters to maximize performance for a specific convolution layer. However, the heterogeneity of CNN models makes it challenging to achieve the global optimum. When an image is fed into a CNN model, it has a high resolution (e.g. $H\times W= 224\times224$) and low channel width (e.g. $C=3$). As the network deepens, the feature map resolution gradually decreases with increasing channel width. The heterogeneity of feature map shapes reflects the variation of loop boundaries in the nested loop. Existing accelerators can only achieve high resource utilization for some layers.

Multi-accelerator systems make it possible to maintain a high utilization rate through the whole network inference. We can deploy different layers onto accelerators that are suitable for corresponding computation patterns. Besides the computation heterogeneity, memory burden is another issue that needs to be addressed. As the depth and width of the network enlarge rapidly, the off-chip DRAM attached to each accelerator may not have sufficient space to buffer all the parameters and intermediate results during the inference. Frequent access to the host memory can lead to high latency. Thus, we need to 1) alleviate the memory burden by reducing the size of parameters and intermediate results, and 2) manage the off-chip DRAM effectively to minimize the host memory accesses. 
\vspace{-0.5cm}
\subsection{Motivation}\label{motivation}

Putting it all together, we realize that mapping DNN workloads on adaptive multi-accelerator systems are a challenging problem with extremely large design space. There are many factors that hinder designers from finding a high-quality mapping strategy. We analyze design choices in the design space.

\noindent \textbf{Choices of accelerator designs \& Workload allocation:} Due to the heterogeneity of DNN layers and various computation patterns that different accelerator designs take, we may select different accelerator designs to configure accelerators in the adaptive multi-accelerator system. Thus, we can allocate the layers in the DNN to those accelerators with their preferred computation patterns. In this way, there could be multiple layers mapped to a set of accelerators with the same design.

\noindent \textbf{Choices of parallelism strategies:} For DNN layers mapped to accelerators with the same design, partitioning each DNN layer with suitable parallelism strategies can further distribute the computation load to these accelerators. This naturally distributes the parameters of each layer and specific strategies are necessary to reduce the memory cost.

Due to the importance and complexity of the above design choices, it is necessary to give a clear formulation at first. We present system formulation in Section \ref{formulation}.

\vspace{-0.2cm}
\section{System Formulation}\label{formulation}

\noindent\textbf{Multi-accelerator system modeling:} We formulate the topology of multi-accelerators as a graph, $G(Acc, BW)$. As shown in Table \ref{tab:notation}, the vertex $Acc_i$ in the graph refers to the accelerator in the system of which the design can be configured adaptively. The weight of the edge $BW_{i,j}$ refers to the bandwidth of the communication between $Acc_i$ and $Acc_j$. This is necessary because of the asymmetrical communication patterns mentioned in Section \ref{Adaptive}. Moreover, each accelerator $Acc_i$ in the system can access the host memory with the bandwidth $BW_{i,host}$. Besides the communication, each accelerator $Acc_i$ is equipped with off-chip DRAM with size $Mem_i$ to buffer the intermediate results and parameters. 

\noindent\textbf{Accelerator designs:} 
For adaptive multi-accelerator systems, there are various existing accelerator designs to choose from. We use a set $Design=\{d_1, ..., d_M\}$ to represent the candidates. Following the concepts in Section \ref{motivation}, we define a set of accelerators with the same design as an accelerator set, denoted by $AccSet$. Correspondingly, we use a map $Config[AccSet_i] = d_i$ to express that the accelerators in $AccSet_i$ are configured to the design $d_i$.
It should be satisfied that 
$AccSet_i \cap AccSet_j = \varnothing$ for any $i\neq j$.
For each accelerator design, similar to H2H\cite{zhang2022h2h}, we use their analytical performance models to evaluate the number of cycles. 



\begin{table}[t]
\caption{Explanations of Main Notations in System Formulation}
\vspace{-0.2cm}
\label{tab:notation}
\begin{tabularx}{0.48\textwidth}{cY}
\toprule
\textbf{Notation} & \textbf{Explanation}\\
\midrule
$Acc_i$& The accelerator that can be configured adaptively\\
\midrule
$BW_{i,j}$& The communication bandwidth between $Acc_i$ and $Acc_j$\\
\midrule
$Mem_i$& The size of off-chip DRAM attached to $Acc_i$\\
\midrule
$d_i$ & The available accelerator design in the adaptive system\\
\midrule
$AccSet_i$ & A set of accelerators with the same design\\
\midrule
$Config$& A map from $AccSet_i$ to the chosen design $d_i$\\
\midrule
$L_i$ & The $i$-th layer of the DNN workload\\
\midrule
$LayerSet_i$& A set of layers mapped to a certain accelerator set\\
\midrule
$Map$& A map from $LayerSet_i$ to $AccSet_i$ \\
\midrule
$ES,SS$ & The sets used to describe parallelism strategies\\
\bottomrule
\end{tabularx}
\vspace{-0.4cm}
\end{table}

\noindent\textbf{DNN workload allocation:} the DNN workload can be represented as a computation graph with a series of layers $\{L_1, ..., L_N\}$ (flattened in topology order). Each layer holds its parameters like $(C_{out}, C_{in}, H, W, K)$ for convolution. In our formulation, a subset of all the layers, $LayerSet_i$, is mapped to a certain accelerator set, $AccSet_i$, which can be expressed as $Map[LayerSet_i]=AccSet_i$.

\noindent\textbf{Parallelism strategies:}
For each $LayerSet$ after DNN workload allocation, 
we further perform parallel strategies to expedite its inference. Specifically, we use two sets $ES$ and $SS$ which describe two different strategies used for partitioning each layer in the layer set. Details will be discussed in Section \ref{layer_parallism}. Note that the chosen parallelism strategies are valid only if the tensor sizes of these partitioned layers do not exceed the DRAM memory space of the corresponding accelerator set.


Table \ref{tab:notation} lists all the notations of our system formulation. MARS can evaluate the whole network inference latency which consists of the latency of each accelerator set and the communication latency between accelerator sets. We use ASTRA-Sim\cite{rashidi2020astra} to simulate communication latency in the system. In addition, we integrate analytical performance models of different accelerator designs into ASTRA-Sim to evaluate the computation cycles. ASTRA-Sim is a simulator featuring collective communication latency estimation for multi-accelerator systems. 
\begin{figure}
    \centering
    \includegraphics[width=0.48\textwidth]{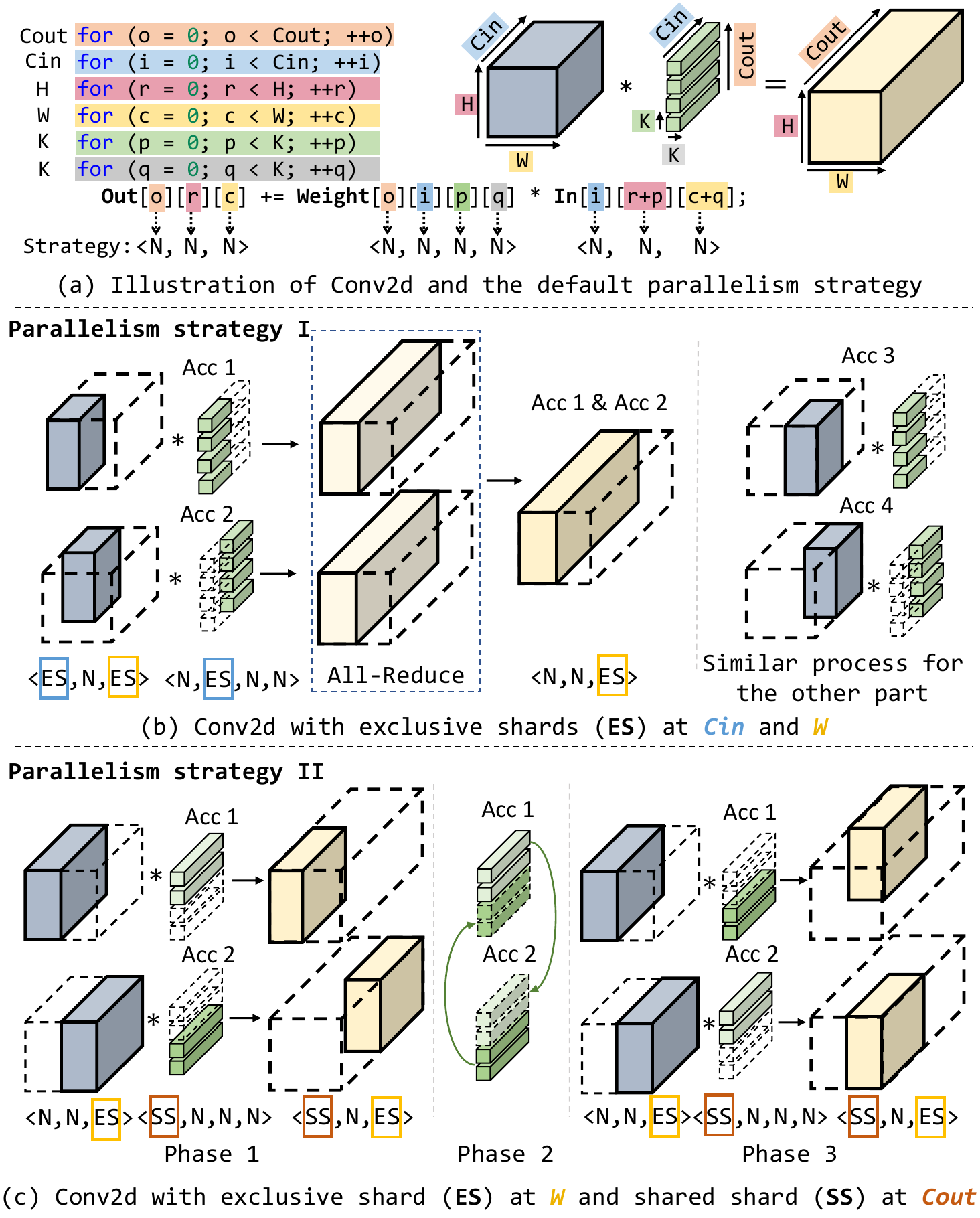}
    \vspace{-0.2cm}
    \caption{Using exclusive/shared shard to parallelize the layer computation}
    \vspace{-0.4cm}
    \label{fig:exclusive_shard}
\end{figure}

\vspace{-0.2cm}
\section{Parallelism Strategies in MARS}\label{layer_parallism} 

\begin{figure*}
    \centering
    \includegraphics[width=0.9\textwidth]{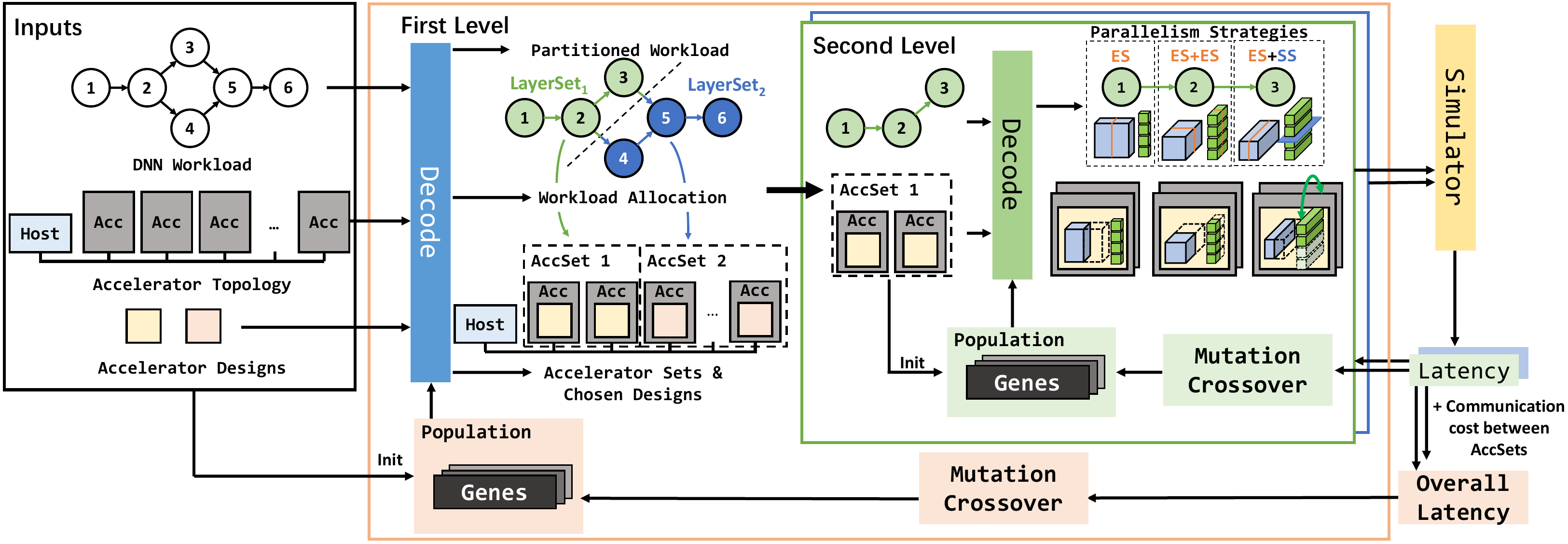}
    \vspace{-0.2cm}
    \caption{The Overview of MARS Mapping Algorithm}
    \vspace{-0.6cm}
    \label{fig:overview}
\end{figure*}

As introduced in our system formulation in Section \ref{formulation}, we can perform different parallelism strategies to layers in a subset $LayerSet$, further exploiting the parallelism and boosting the performance.  
There have been some previous works applying parallelism strategies on GPU clusters\cite{zheng2022alpa}, and chiplets\cite{tan2021nn}. And we generalize them and include the additional \textbf{SS} strategy for accelerators. The computation-intensive layers like convolution in a DNN workload can usually be represented as a nested loop, as shown in Fig. \ref{fig:exclusive_shard}(a). In this example, it takes the input feature map \textit{In} with shape $(C_{in}, H, W)$ and \textit{Weight} with shape $(C_{out}, C_{in}, K, K)$ as input, and produces the output feature map \textit{Out} with shape $(C_{out}, H, W)$. The essence of layer-level parallelism is partitioning the nested loop along several dimensions. Then the tensors involved in these dimensions will be partitioned into different shards. We can distribute these shards to accelerators in the system and let them perform the computation in parallel. Figure. \ref{fig:exclusive_shard} illustrates the idea of our parallelism strategies. By default, As shown in Fig. \ref{fig:exclusive_shard}(a), tensors are not partitioned and we annotate each dimension with \textbf{N} to represent the nested loop is not partitioned at the corresponding dimension. Figure \ref{fig:exclusive_shard}(b) and 2(c) presents two examples of our parallelism strategies, using different ways to generate shards.
Specifically, we classify tensor shards into two categories: exclusive shards and shared shards. And we represent the parallelism strategy by annotating corresponding dimensions with \textbf{ES} and \textbf{SS}.
\begin{itemize}[leftmargin=10pt, rightmargin=0cm]
    \item \textbf{Exclusive Shards (ES)} of a tensor are held exclusively by accelerators participating in the parallel computation of a layer. Note that a tensor can be partitioned into exclusive shards along multiple specified dimensions.
    \item \textbf{Shared Shards (SS)} of a tensor can be shared among accelerators. At the beginning of the computation, each accelerator holds one shared shard of the tensor. After the computation, the shared shards are transmitted to the neighbor accelerator, which will form a ring communication logically. Such computation and communication processes will alternate until accelerators finish computation using all shared shards of the tensor. 
\end{itemize}



In Fig.\ref{fig:exclusive_shard}(b), \textit{In} and \textit{Out} are split along the $W$-dimension (marked in blue) into two exclusive shards. \textit{In} and \textit{Weight} are split along the $C_{in}$-dimension (marked in yellow) into two exclusive shards. By performing the \textbf{ES} strategy along these two dimensions, the computation workload is partitioned into four equal parts, each of which can be executed on four accelerators (Acc1-Acc4). In this way, each accelerator holds a quarter of the input feature map and half of the weight. As shown in Fig.\ref{fig:exclusive_shard}(b), because Acc1 and Acc2 generate the same left half of the output feature map \textit{Out} (i.e., left side of the cuboid), the results need to be summed up using the collective communication primitive, All-Reduce. The same process is performed on Acc3 and Acc4 to compute the right half of \textit{Out}. After that, \textit{Out} is equally split into two parts along the partitioned dimension, $W$, stored in Acc1\&Acc2 and Acc3\&Acc4 respectively. 

In addition to exclusive shards, parameters can further be partitioned into shared shards. Fig. \ref{fig:exclusive_shard}(c) shows how exclusive shards and shared shards coexist in multi-accelerator systems. $In$ and $Out$ are split into exclusive shards along the $W$-dimension. $Weight$ and $Out$ are split into shared shards along the $C_{out}$-dimension. The workload can be executed on two accelerators because of the sharing of parameters. In phase 1, Acc1 and Acc2 hold exclusive shards of $In$ and shared shards of $Weight$, and generate the corresponding quarters of $Out$. Then in phase 2, Acc1 and Acc2 transmit the shared shards of $Weight$ they hold to each other. In phase 3, the two accelerators compute the remaining part of $Out$ using the received shards of $Weight$. The \textbf{SS} strategy can be useful due to the communication heterogeneity in adaptive multi-accelerator systems, as introduced in Section \ref{Adaptive}. It can leverage the low-latency communication between accelerators within a group or a rack and hence relieve the data bus congestion to/from the host. 

We represent the parallelism strategies formally according to the dimensions which are partitioned. For example, the strategies can be expressed as two sets, i.e., $ES=\{C_{in}, W\}, SS=\varnothing$ in Fig. \ref{fig:exclusive_shard}(b) and $ES=\{W\}, SS=\{C_{out}\}$ in Fig. \ref{fig:exclusive_shard}(c), respectively. Because of the coexistence of exclusive shards and shared shards in multi-level nested loops, we need to select suitable parallelism strategies for all the dimensions from a wide range of candidates. When applying exclusive shards on two dimensions of the convolution layers, there are $C_{6}^{2}=15$ choices. In addition, when applying shared shards on one certain dimension, the number of choices increases to $C_{6}^{2}\cdot 6=90$. They will incur different communication volumes and computation patterns. For the DNN workloads, the number of design points in the design space increases exponentially with respect to the depth of the network. To efficiently search the large design space to find a design point with low system latency, we propose the mapping algorithm in MARS.

\vspace{-0.2cm}
\section{MARS Mapping Algorithm}\label{mapping_algorithm}
Figure \ref{fig:overview} presents the overview of MARS mapping algorithm, using a two-level genetic algorithm to explore the design space. Genetic algorithms are widely used in deep learning kernel generation\cite{vasilache2018tensor} and optimization on multi-core accelerators\cite{kao2022magma}. We mainly focus on our improvements to adapt genetic algorithms to the mapping problem.

As discussed in Section \ref{formulation}, The accelerator configurations and workload allocation may deeply influence the optimal parallelism strategy: the layer with the same parallelism strategy can show a large performance gap when the accelerator design changes. Simply tuning them in one pass of the search is easy to fall into local optimums. To make the problem more solvable, we adopt the idea of divide and conquer by dividing the problem into two levels. The first-level genetic algorithm (the pink box in Fig. \ref{fig:overview}) aims at finding the minimum overall latency for the multi-accelerator system. At the first level, $AccSet$ with configured designs, and the layers mapped to them, $LayerSet$, are decided. With these factors fixed, the mapping problem is further divided into several sub-problems: For each layer set $LayerSet_i$ mapped to an accelerator set $AccSet_i$, what parallelism strategies should be applied to the layers in these layer sets? To solve these sub-problems, the second-level genetic algorithm (green and blue boxes in Fig. \ref{fig:overview}) is responsible to find the proper parallelism strategy for each layer to minimize the latency on the given accelerator set, considering both computation and communication costs. The population is iteratively updated in the mutation and crossover phase according to the latency evaluated by the simulator. After several iterations, the minimum latencies of accelerator sets will be aggregated to obtain the overall latency with extra communication costs between the sets. The overall latency will work as the fitness function to influence the succeeding mutation and crossover of the first-level genetic algorithm.

We use several heuristics to prune the search space. The topology of multi-accelerator systems is formulated as a graph, $G(Acc, BW)$, which has been introduced in Section \ref{formulation}. MARS iteratively removes the edge with the lowest bandwidth in $G(Acc, BW)$. This will produce several connected sub-graphs, which are regarded as candidates of $AccSet$. This strategy can help generate $AccSet$ with minimal communication bottlenecks. In the decode step in Fig.\ref{fig:overview}, the candidate of $AccSet$ with the highest gene value will be chosen. The accelerator design of each $AccSet$ is decided through the gene value of each design. MARS profiles the performance of accelerator designs on the layers of the DNN workload according to analytical models before the search. The gene value of these designs at the first generation is initialized according to the normalized performance. This means the design with higher computation ability is most likely to be chosen at the beginning of the search.
As for layer sets, to avoid frequent communication between different accelerator sets, we limit that each accelerator set is only mapped with a continuous series of layers in topology order. 
Then MARS will call the second-level genetic algorithm to optimize parallelism strategies over the accelerator set and the layer set.
At the second level, the parallelism strategies of each layer are decided. For layer $L$, MARS uses individual genes to decide the $ES$ and $SS$ sets of $L$ respectively. It prioritizes parallelism at the dimensions with higher gene values. 
\vspace{-0.2cm}
\section{Evaluation}
\subsection{Experiment Setup}\label{settings}
\noindent\textbf{Hardware Platform Modeling:} We model an adaptive multi-accelerator system in modified ASTRA-Sim\cite{rashidi2020astra}. The system topology is modeled based on the interconnection of F1 instances, as illustrated in Fig. \ref{fig:F1}. The system consists of eight accelerators separated into two groups. 
We set the communication bandwidth between accelerators in the same group to 8Gbps. The accelerator-to-host bandwidth is set to 2Gbps to simulate the high latency of accessing host memory. The size of off-chip DRAM on accelerators is set to 1GB.

\noindent\textbf{Accelerator Designs:} For the accelerator designs, we use three kinds of CNN accelerators on FPGA with their performance models. To make their theoretical performance comparable, we set the clock frequency to 200MHz uniformly and use design parameters with similar numbers of PEs. The detailed setting can be found in Table \ref{tab:accelerators}.

\begin{table}
\caption{Available Accelerator Designs}
\vspace{-0.2cm}
\label{tab:accelerators}
\begin{tabular}{m{0.8em} m{1cm} m{1cm} m{0.8cm} m{3.6cm}}
\toprule
& Design & Freq(MHz) &  \#PEs & Design Parameters \\
\midrule
1 & SuperLIP\cite{jiang2019achieving} & 200 & 438 & $T_m, T_n, T_r, T_c:64,7,7,14$ \\
\midrule
2 & \cite{wei2017automated} & 200 & 572& $row,col,vec: 11,13,8$ \\
\midrule
3 & \cite{lu2017evaluating} & 200 & 576 & $n, P_n, P_m:6,2,8$
\\
\bottomrule
\end{tabular}
\vspace{-0.6cm}
\end{table}
\noindent\textbf{Models:} We use several representative CNN models as benchmarks for evaluation, including AlexNet, VGG, ResNet, and WideResNet. 
For some models, we use multiple settings with different parameter sizes and FLOPs.
\begin{table*}[!t]
\caption{Latency comparison between baseline and MARS}
\label{tab:comparison}
\vspace{-0.2cm}
\begin{tabularx}{\linewidth}{C{1.2cm}C{0.8cm}C{0.8cm}C{0.9cm}lC{2cm}Y}
\toprule
\textbf{Model} & \textbf{\#Convs}&\textbf{\#Params} & \textbf{FLOPs} & \textbf{Algorithm} & \textbf{Latency}/ms &  Mapping found by MARS\\
\midrule
\multirow{2}{5em}{AlexNet} &\multirow{2}{5em}{5} & \multirow{2}{5em}{61.1M} & \multirow{2}{5em}{727M} & Baseline & 0.832 & \multirow{2}{35em}{Conv1-2→4$\times$Design 1;\ \quad Conv1:$ES=\{H, W\},SS=\varnothing$\newline Conv3-5→4$\times$Design 3; \quad Conv5:$ES=\{H, C_{Out}\},SS=\varnothing$}\\ 
&& && MARS & 0.748(-10.1\%) & \\
\midrule
\multirow{3}{5em}{VGG16} &\multirow{3}{5em}{13} & \multirow{3}{5em}{138M} & \multirow{3}{5em}{15.5G} & Baseline & 20.6 & \multirow{2}{35em}{Conv1-7→4$\times$Design 1;\ \quad Conv1:$ES=\{H, W\},SS=\varnothing$\newline Conv8-10→2$\times$Design 2;\quad Conv8:$ES=\{H\},SS=\{C_{Out}\}$\newline Conv11-13→2$\times$Design 3;\  Conv13:$ES=\{C_{Out}\},SS=\{H\}$}\\ 
&& && MARS & 14.9(-27.7\%) & \\
&&&&&&\\
\midrule
\multirow{2}{5em}{ResNet34} &\multirow{2}{5em}{33} & \multirow{2}{5em}{21.8M} & \multirow{2}{5em}{3.68G} & Baseline & 4.43 & \multirow{2}{35em}{Conv1-15→4$\times$Design 1;\ \quad Conv2:$ES=\{H, W\},SS=\varnothing$\newline Conv16-33→4$\times$Design 3;\quad Conv29:$ES=\{C_{In}, C_{Out}\},SS=\varnothing$}\\ 
&& && MARS & 2.76(-37.7\%) & \\
\midrule
\multirow{3}{5em}{ResNet101} &\multirow{3}{5em}{100} & \multirow{3}{5em}{44.55M} & \multirow{3}{5em}{7.85G} & Baseline &14.9 & \multirow{3}{35em}{Conv1-22→2$\times$Design 1;\quad\quad Conv2:$ES=\{H\},SS=\varnothing$\newline Conv23-79→4$\times$Design 2;\ \quad Conv24:$ES=\{H, W\},SS=\{C_{Out}\}$\newline
Conv80-100→2$\times$Design 2;\quad Conv99:$ES=\{C_{Out}\},SS=\{W\}$
}\\ 
&& && MARS & 7.95(-46.6\%) & \\
&&&&&& \\
\midrule
\multirow{3}{5em}{WRN-50-2} &\multirow{3}{5em}{49} & \multirow{3}{5em}{68.8M} & \multirow{3}{5em}{11.4G} & Baseline & 16.7 & \multirow{3}{35em}{Conv1-17→2$\times$Design 1;\ \quad Conv3:$ES=\{H\},SS=\varnothing$\newline Conv18-29→2$\times$Design 2;\quad Conv18:$ES=\{C_{Out}\},SS=\{H\}$\newline
Conv30-49→4$\times$Design 2;\quad Conv48:$ES=\{C_{In},C_{Out}\},SS=\varnothing$}\\ 
&& && MARS & 10.1(-39.5\%) & \\
&&&&&& \\
\bottomrule
\end{tabularx}
\vspace{-0.5cm}
\end{table*}

\noindent\textbf{Baseline:} To the best of our knowledge, there are no existing mapping algorithms that take both multi-level parallelism and adaptive multi-accelerator systems into consideration. To show the ability of MARS, we extend the computation-prioritized mapping algorithm from \cite{kwon2021heterogeneous} with parallelism strategies integrated. The baseline uses fixed two accelerator sets which are the same as two groups in the system topology. This is reasonable to avoid high communication latency across groups. And it allocates half of the layers to each accelerator set and chooses the accelerator design with the lowest computation latency. About the parallelism strategies, each layer is partitioned with ES along the longest two dimensions.

\vspace{-0.2cm}
\subsection{Performance Analysis}
Following the experiment setup, we test the latency of the baseline mapping algorithm and MARS mapping algorithm. We list the parameters of the models, the overall latency, together with the accelerator sets, workload allocation, and parallelism strategies of representative layers in each layer set found by MARS in Table \ref{tab:comparison}.

As shown in the table, MARS outperforms the baseline for all models. The latency reduction ranges from 10.1\% to 46.6\% (32.2\% on average). The larger design space enables MARS to find better solutions compared to the baseline. Some patterns are shown in the mappings found by MARS. The first few layers of these models are always mapped to accelerator sets configured with Design 1(SuperLIP). The reason is that the first few layers usually have larger resolutions and fewer channels. Other designs suffer from low hardware utilization because the shape of the layer cannot saturate the PEs in the architecture, while the design parameter of SuperLIP($T_n=7$) can achieve relatively high utilization. MARS tends to partition these layers along $H/W$-dimension. We can also find that design 3 does not show up in ResNet101 and WRN-50-2. This is because design 3 is an accelerator based on Winograd algorithm, which makes it impossible to effectively handle $1\times 1$ convolution in the bottleneck block of these models. Because $C_{In}$ and $C_{Out}$ enlarge rapidly when the network goes deeper, MARS is more likely to partition these layers along $C_{In}/C_{Out}$-dimension for parallelism. This emphasizes the importance to select accelerator designs and parallelism strategies based on the computation patterns of layers.

\vspace{-0.2cm}
\subsection{Comparison with H2H}
\vspace{-0.1cm}
H2H\cite{zhang2022h2h} focuses on mapping heterogeneous models to heterogeneous multi-accelerators with fixed accelerator designs. Though MARS and H2H have different problem formulations, we still compare their performance with heterogeneous DNN models. We reuse performance models of convolution accelerators used in H2H and model the cloud-scale multi-FPGA systems in ASTRA-Sim following the 5-level bandwidth settings of H2H. For heterogeneous accelerator designs, 
we assume that members in the accelerator set stall until the slowest accelerator finishes computing. 
Then we evaluate the latencies in milliseconds of MARS and H2H on two ResNet-based heterogeneous models. The results are shown in Table \ref{tab:H2H}. 
We can see that MARS achieves lower latency than H2H on both models when limited to the same bandwidth (59.4\% reduction on average). When the bandwidth is extremely low, MARS tends to partition convolution layers along $H/W$-dimension, which requires low communication cost.
\vspace{-0.3cm}
\begin{table}
\caption{Comparison of Latency ($\mathrm{ms}$) with H2H}
\label{tab:H2H}
\vspace{-0.2cm}
\begin{tabularx}{\linewidth}{YC{0.8cm}C{1.7cm}C{0.8cm}C{1.7cm}}
\toprule
\multirow{2}{5em}{\textbf{Bandwidth}}  & \multicolumn{2}{c}{\textbf{CASIA-SURF}\cite{zhang2020casia}} & \multicolumn{2}{c}{\textbf{FaceBag}\cite{shen2019facebagnet}}\\

& H2H & MARS & H2H & MARS\\
\midrule
\scriptsize Low-(1Gbps) & 360.0 & 124.6(-65.4\%) & 520.0 & 237.4(-54.3\%) \\
\midrule
\scriptsize Low(1.2Gbps) & 340.0 & 120.3(-64.6\%) & 450.0 & 224.6(-50.1\%) \\
\midrule
\scriptsize Mid-(2Gbps) & 260.0 & 100.9(-61.2\%) & 320.0 & 159.4(-50.2\%) \\
\midrule
\scriptsize Mid(4Gbps) & 230.0 & 74.3(-67.7\%) & 230.0 & 112.1(-51.3\%) \\
\midrule
\scriptsize High(10Gbps) & 180.0 & 46.8(-74.0\%) & 170.0 & 76.5(-55.0\%) \\
\bottomrule
\end{tabularx}
\vspace{-0.7cm}
\end{table}

\section{Conclusion}
\vspace{-0.12cm}
In this paper, we propose MARS, a mapping framework aiming at exploiting multi-level parallelism on adaptive multi-accelerator systems. We formulate design space including the choices of accelerator design, workload allocation, and parallelism strategies. A two-level genetic algorithm with heuristics is used to perform design space exploration. The mapping algorithm shows significant latency reduction compared to the baseline mapping algorithm and the state-of-the-art method. 

\vspace{-0.15cm}
\section{Acknowledgements}
\vspace{-0.15cm}
This work is partially sponsored by the National Natural Science Foundation of China (62102249, 62232015) and Shanghai Pujiang Program (21PJ1408200).



 

\vspace{-0.2cm}
\bibliographystyle{unsrt}
\bibliography{ref}

\begin{thebibliography}{10}

\bibitem{he2016deep}
Kaiming He et~al.
\newblock Deep residual learning for image recognition.
\newblock In {\em CVPR}, pages 770--778, 2016.

\bibitem{vaswani2017attention}
Ashish Vaswani et~al.
\newblock Attention is all you need.
\newblock {\em Advances in neural information processing systems}, 30, 2017.

\bibitem{cheng2016wide}
Heng-Tze Cheng et~al.
\newblock Wide \& deep learning for recommender systems.
\newblock In {\em Proceedings of the 1st workshop on deep learning for
  recommender systems}, pages 7--10, 2016.

\bibitem{fowers2018configurable}
Jeremy Fowers et~al.
\newblock A configurable cloud-scale dnn processor for real-time ai.
\newblock In {\em ISCA}, pages 1--14. IEEE, 2018.

\bibitem{zhang2019efficient}
Wentai Zhang et~al.
\newblock An efficient mapping approach to large-scale dnns on multi-fpga
  architectures.
\newblock In {\em DATE}, pages 1241--1244. IEEE, 2019.

\bibitem{kwon2021heterogeneous}
Hyoukjun Kwon et~al.
\newblock Heterogeneous dataflow accelerators for multi-dnn workloads.
\newblock In {\em HPCA}, pages 71--83. IEEE, 2021.

\bibitem{zhang2022h2h}
Xinyi Zhang et~al.
\newblock H2h: Heterogeneous model to heterogeneous system mapping with
  computation and communication awareness.
\newblock {\em DAC}, 2022.

\bibitem{peer2peer}
deeppat kristopk.
\newblock How to use the pcie peer-2-peer version 1.0, 2021.

\bibitem{rashidi2020astra}
Saeed Rashidi et~al.
\newblock Astra-sim: Enabling sw/hw co-design exploration for distributed dl
  training platforms.
\newblock In {\em ISPASS}, pages 81--92, 2020.

\bibitem{zheng2022alpa}
Lianmin Zheng et~al.
\newblock Alpa: Automating inter-and intra-operator parallelism for distributed
  deep learning.
\newblock {\em OSDI}, 2022.

\bibitem{tan2021nn}
Zhanhong Tan et~al.
\newblock Nn-baton: Dnn workload orchestration and chiplet granularity
  exploration for multichip accelerators.
\newblock In {\em ISCA}, pages 1013--1026. IEEE, 2021.

\bibitem{vasilache2018tensor}
Nicolas Vasilache et~al.
\newblock Tensor comprehensions: Framework-agnostic high-performance machine
  learning abstractions.
\newblock {\em arXiv preprint arXiv:1802.04730}, 2018.

\bibitem{kao2022magma}
Sheng-Chun Kao et~al.
\newblock Magma: An optimization framework for mapping multiple dnns on
  multiple accelerator cores.
\newblock In {\em HPCA}, pages 814--830. IEEE, 2022.

\bibitem{jiang2019achieving}
Weiwen Jiang et~al.
\newblock Achieving super-linear speedup across multi-fpga for real-time dnn
  inference.
\newblock {\em ACM TECS}, 18(5s):1--23, 2019.

\bibitem{wei2017automated}
Xuechao Wei et~al.
\newblock Automated systolic array architecture synthesis for high throughput
  cnn inference on fpgas.
\newblock In {\em DAC}, pages 1--6, 2017.

\bibitem{lu2017evaluating}
Liqiang Lu et~al.
\newblock Evaluating fast algorithms for convolutional neural networks on
  fpgas.
\newblock In {\em FCCM}, pages 101--108. IEEE, 2017.

\bibitem{zhang2020casia}
Shifeng Zhang et~al.
\newblock Casia-surf: A large-scale multi-modal benchmark for face
  anti-spoofing.
\newblock {\em IEEE TBIOM}, 2(2):182--193, 2020.

\bibitem{shen2019facebagnet}
Tao Shen et~al.
\newblock Facebagnet: Bag-of-local-features model for multi-modal face
  anti-spoofing.
\newblock In {\em Proceedings of CVPR Workshops}, 2019.

\end{thebibliography}
\end{document}